\documentclass[10pt,draftcls,onecolumn]{IEEEtran}
\ifCLASSINFOpdf
\else
   \usepackage[dvips]{graphicx}
\fi
\usepackage{url}
\usepackage{graphicx}
\usepackage{amsmath}
\usepackage{comment}
\usepackage[utf8]{inputenc}
\usepackage[english]{babel}
\usepackage{color}
\usepackage{framed}
\usepackage{amsmath,amssymb,amsfonts}

\usepackage{graphicx}
\usepackage{upgreek}
\usepackage{textcomp}

\usepackage{url}

\usepackage{cite}

\usepackage[dvipsnames]{xcolor}
\usepackage{amsmath}
\usepackage{amsthm}
\usepackage{dsfont}
\usepackage{thmtools}
\usepackage{amssymb}

\def\bSigma{\boldsymbol{\Sigma}}

\def\mbb{\mathbf{b}}

\def\mbN{\mathbf{N}}

\def\mbQ{\mathbf{Q}}
\def\mbR{\mathbf{R}}

\def\mbU{\mathbf{U}}
\def\mbV{\mathbf{V}}

\def\mbX{\mathbf{X}}
\def\mbY{\mathbf{Y}}
\def\mbZ{\mathbf{Z}}

\newtheorem{theorem}{Theorem}

\newtheorem{lemma}{Lemma}

\theoremstyle{definition}
\newtheorem{definition}{Definition}

\usepackage{bbold}
\usepackage{ifpdf}
\usepackage{times}
\usepackage{layout}
\usepackage{float}
\usepackage{afterpage}
\usepackage{amsmath}
\usepackage{amstext}
\usepackage{amssymb,bm}
\usepackage{latexsym}
\usepackage{color}
\usepackage{flexisym}
\usepackage{kantlipsum}
\usepackage{graphicx}
\usepackage{amsmath}
\usepackage{amsthm}
\usepackage{graphicx}

\usepackage[caption=false,font=footnotesize]{subfig}

\usepackage{booktabs}
\usepackage{multicol}

\usepackage{enumitem}

\usepackage[switch]{lineno}
\usepackage[named]{algo}
\usepackage[noend]{algpseudocode}
\usepackage{algorithm}

\makeatletter
\newcommand*{\rom}[1]{\expandafter\@slowromancap\romannumeral #1@}
\makeatother

\begin{document}
\setlength{\abovedisplayskip}{3pt}
\setlength{\belowdisplayskip}{3pt}

\title{Automotive Radar Sensing with Sparse Linear Arrays Using One-Bit Hankel Matrix Completion
}

\author{
Arian Eamaz, Farhang Yeganegi, Yunqiao Hu, Shunqiao Sun, and Mojtaba Soltanalian  

\vspace{-20pt}
\thanks{Arian Eamaz, Farhang Yeganegi and Mojtaba Soltanalian are with the ECE Department, University of Illinois at Chicago, Chicago, IL 60607 USA.
}
\thanks{Yunqiao Hu and Shunqiao Sun are with the Department of Electrical and Computer Engineering, University of Alabama, Tuscaloosa, AL 35487 USA.}
\thanks{This work was supported in part by the National Science Foundation Grants CCF-1704401 and CCF-2153386. (\emph{Corresponding author: Arian Eamaz})}
}
\markboth{
}
{Shell \MakeLowercase{\textit{et al.}}: Bare Demo of IEEEtran.cls for IEEE Journals}
\maketitle

\begin{abstract}
The design of sparse linear arrays has proven instrumental in the implementation of cost-effective and efficient automotive radar systems for high-resolution imaging. This paper investigates the impact of coarse quantization on measurements obtained from such arrays. To recover azimuth angles from quantized measurements, we leverage the low-rank properties of the constructed Hankel matrix. In particular, by addressing the one-bit Hankel matrix completion problem through a developed singular value thresholding algorithm, our proposed approach accurately estimates the azimuth angles of interest. We provide comprehensive insights into recovery performance and the required number of one-bit samples. The effectiveness of our proposed scheme is underscored by numerical results, demonstrating successful reconstruction using only one-bit data.
\end{abstract}


\setlength{\abovedisplayskip}{3pt}
\setlength{\belowdisplayskip}{3pt}

\section{Introduction}
\label{intro}
 Millimeter wave (mmWave) automotive radars are highly reliable in various weather environments, with antennas that can be fit into a small area to provide high angular resolution. Benefiting from multiple-input multiple-output (MIMO) radar technology, mmWave radars can synthesize virtual arrays with large aperture sizes using a small number of transmit and receive antennas\cite{sun2020mimo}. To further reduce hardware cost, sparse arrays synthesized by MIMO radar technology have been widely adopted in automotive radar \cite{sun2020sparse, sun2023fast}.
Investigating single-snapshot Direction-of-Arrival (DoA) estimation with sparse arrays is crucial, particularly in dynamic automotive contexts where, often, only limited or single radar snapshots are available. 
The challenges associated with single-snapshot DoA with sparse arrays are the high sidelobes and the reduction of signal-to-noise ratio (SNR), both of which may cause errors and ambiguity in estimation\cite{sun2020sparse}. One potential solution is interpolating the missing elements in the sparse array using techniques such as matrix completion\cite{sun20214d,sun2020sparse,zhang2021enhanced}, followed by standard DoA estimation algorithms like MUSIC and ESPRIT. Matrix completion approach exploits the low-rank property of the Hankel matrix formulated by array received signals, and completes the missing elements using iterative algorithms~\cite{chen2013spectral,fazel2013hankel}.

Quantization, a crucial step in digital signal processing, transforms continuous signals into discrete representations. Traditional high-resolution quantization often requires numerous quantization levels, leading to higher power consumption, increased manufacturing costs, and reduced analog-to-digital converter (ADC) sampling rates. In the pursuit of alternative systems, researchers have explored reduced quantization bits, including one-bit quantization, where signals are compared with a fixed threshold at ADCs, yielding binary outputs \cite{ boufounos2015quantization}. This one-bit approach enables high-rate sampling while reducing implementation costs and energy consumption compared to multi-bit ADCs. One-bit ADCs find valuable applications in MIMO systems \cite{kong2018multipair, mezghani2018blind,deng2022one}, channel estimation \cite{li2017channel}, target detection \cite{ameri2019one,ameri20191one}, and array signal processing \cite{liu2017one}.

Scalar quantization with a dithering scheme involves adding random dither to an input signal before quantization. This well-established technique is recognized for enhancing resolution and signal reconstructions in practical applications and significantly reducing quantization noise in theoretical contexts \cite{gray1993dithered,vanderkooy1987dither,feuillen2019quantity}. Unlike sigma-delta quantization, scalar quantization operates as a memoryless scheme, requiring no feedback or update process, a characteristic widely acknowledged and referenced in the literature \cite{wagdy1989validity,carbone1997quantitative,ali202012}.

In this paper, we employ coarse quantization on the non-missing entries of the acquired Hankel matrix from a sparse linear array. Subsequently, we fill the missing parts of the matrix using the well-established iterative approach known as singular value thresholding (SVT). This process results in utilizing exclusively one-bit data for the subsequent reconstruction algorithm. 
It has been demonstrated that when the scalar parameter of uniform dithers is designed to dominate the dynamic range of measurements, multi-bit scalar quantization simplifies to a one-bit comparator\cite{thrampoulidis2020generalized,eamaz2022uno}. Leveraging this observation along with the properties of uniform dithering, particularly in canceling quantization effects in the expectation, we aim to establish theoretical guarantees for quantized matrix completion.

In the numerical results, we demonstrate that employing one-bit quantization with an appropriate dithering scheme allows the generated one-bit dither samples to be effectively utilized in the reconstruction step. This approach yields meaningful results, such as accurately detecting the two azimuth locations of targets with high resolution.

\underline{\emph{Notation}}: Throughout this paper, we use bold lowercase and bold uppercase letters for vectors and matrices, respectively. We represent a vector $\mathbf{x}$ and a matrix $\mbX$ in terms of their elements as $\mathbf{x}=[x_{i}]$ and $\mathbf{X}=[X_{i,j}]$, respectively. $\mathbb{C}$ represents the set of complex numbers. $(\cdot)^{\top}$ denotes the vector/matrix transpose. Given a scalar $x$, we define the operator $(x)^{+}$ as $\max\left\{x,0\right\}$.
The nuclear norm of a matrix $\mbX\in \mathbb{C}^{n_1\times n_2}$ is denoted by $\left\|\mbX\right\|_{\star}=\sum^{r}_{i=1}\sigma_{i}$ where $r$ and $\left\{\sigma_{i}\right\}^{r}_{i=1}$ are the rank and singular values of $\mbX$, respectively. The Frobenius norm of a matrix $\mathbf{X}\in\mathbb{C}^{n_1\times n_2}$ is defined as $\|\mathbf{X}\|_{\mathrm{F}}=\sqrt{\sum^{n_1}_{r=1}\sum^{n_2}_{s=1}\left|x_{rs}\right|^{2}}$, where $x_{rs}$ is the $(r,s)$-th entry of $\mathbf{X}$. In the real case, we also define $\|\mbX\|_{\mathrm{max}}=\sup_{i,j}|X_{i,j}|$. The $\ell_{p}$-norm of a vector $\mathbf{x}$ is $\|\mathbf{x}\|_{p}=\left(\sum_{i}x^{p}_{i}\right)^{1/p}$. The operator $\operatorname{diag}\left\{\mathbf{b}\right\}$ denotes a diagonal matrix with $\{b_{i}\}$ as its diagonal elements. The Hadamard (element-wise) product is $\odot$.
The notation $x \sim \mathcal{U}_{[a,b]}$ means a random variable drawn from the uniform distribution over the interval $[a,b]$. If there exists a $c>0$ such that $a\leq c b$ (resp. $a\geq c b$) for two quantities
$a$ and $b$, we have $a\lesssim b$ (resp. $a\gtrsim b$). The uniform quantizer applied to a fixed value $x$ is defined as $\mathcal{Q}_{_{\Delta}}\left(x\right)=\Delta\left(\left\lfloor\frac{x+\tau}{\Delta}\right\rfloor+\frac{1}{2}\right)$ with $\tau\sim \mathcal{U}_{\left[-\frac{\Delta}{2},\frac{\Delta}{2}\right]}$.
\section{High-Resolution Imaging Radar System With One-Bit Measurements}
In this section, we begin by introducing the Hankel matrix completion problem tailored for the sparse linear array system. Subsequently, we delve into the application of dithered one-bit quantization to the resulting Hankel matrix and proceed to recover the signal by addressing the \emph{one-bit Hankel matrix completion} problem. A refined version of the SVT algorithm is presented for recovering the matrix from incomplete observed one-bit data, and the theoretical guarantees of the problem are discussed.
\subsection{Radar Sensing With Sparse Linear Array}
A sparse linear array's antenna positions can be considered a subset of a uniform linear array (ULA) antenna positions. Without loss of generality, let the antenna positions of an $M$-element ULA be $\left \{ kd \right \}$, $k=0,1,\cdots,M-1$, where $d=\frac{\lambda }{2}$ is the element spacing with wavelength $\lambda$. Assume there are $P$ uncorrelated far-field target sources in the same range Doppler bin. The impinging signals on the ULA antennas are corrupted by additive white Gaussian Noise with variance of $\sigma^2$. For the single-snapshot case, only the data collected from a single instance in time is available, resulting in the discrete representation of the received signal from a ULA as
\begin{align}
\mathbf{x} =  \mathbf{A}\mathbf{s} +  \mathbf{n},
\label{signal_model}
\end{align}
where 
\begin{equation}
\begin{aligned}
\mathbf{x} &= \left[x_{1},x_{2},\dots,x_{M}\right]^{\top},\\  \mathbf{A} &= \left[\mathbf{a}\left(\theta_{1}\right),\mathbf{a}\left(\theta_{2}\right)\dots,\mathbf{a}\left(\theta_{P}\right)\right]^{\top},
\end{aligned}
\end{equation}
with 
\begin{align}
\mathbf{a}\left(\theta_{k}\right) = \left [1,e^{j2\pi \frac{d\sin\left(\theta_{k}\right)}{\lambda}},\dots , e^{j2\pi \frac{\left (M-1\right)d\sin \left(\theta_{k}\right)}{\lambda}}\right]^{\top},
\end{align}
for $k = 1,\cdots,P$ and $\mathbf{n} = \left[n_{1},n_{2},\dots,n_{M}\right]^{\top}$. Then, a Hankel matrix denoted as ${\mathcal{H}}\left( {{{\bf{x}}}} \right) \in {\mathbb C}^{n_1 \times n_2}$
where ${n_{1} + n_{2}} = {M} + 1$,  can be constructed from $\mathbf{x}$ \cite{heinig2011fast}. The Hankel matrix ${\mathcal{H}}\left( {\bf{x}} \right)$ admits a Vandermonde decomposition  structure \cite{ying2018vandermonde,sun20214d,cai2019fast}, i.e.,
\begin{align}
{\mathcal{H}}\left({{{\bf{x}}}}\right) = {{\bf{V}}_1}{\bf{\Sigma V}}_2^{\top}, 
\end{align}
where ${{\bf{V}}_{1}} = \left[ {{\bf{v}}_{1}\left( {{\theta_1}} \right), \cdots ,{\bf{v}}_{1}\left( {{\theta _P}} \right)} \right]$, ${{\bf{V}}_{2}} = \left[ {{\bf{v}}_{2}\left( {{\theta _1}} \right), \cdots ,{\bf{v}}_{2}\left( {{\theta _P}} \right)} \right]$ with
\begin{align}
{\bf{v}}_{1}\left( {{\theta _k}} \right) &= {\left[ {1,{e^{j2\pi \frac{{d\sin \left( {{\theta _k}} \right)}}{\lambda }}},\cdots, {e^{j2\pi \frac{{\left( {{n_1} - 1} \right)d\sin \left( {{\theta _k}} \right)}}{\lambda }}}} \right]^{\top}}, \\
{\bf{v}}_{2}\left( {{\theta_k}} \right) &= {\left[ {1,{e^{j2\pi \frac{{d\sin \left( {{\theta _k}} \right)}}{\lambda }}},\cdots, {e^{j2\pi \frac{{\left( {n_2 - 1} \right)d\sin \left( {{\theta _k}} \right)}}{\lambda }}}} \right]^{\top}},
\end{align}
and ${\bf{\Sigma }} = {\rm{diag}}\left( {\left[ {{\sigma_1}, {\sigma_2},\cdots ,{\sigma _P}} \right]} \right)$. Assuming that $P\le\min\left ( n_{1},n_{2}\right)$, and both ${\bf{V}}_{1}$ and ${\bf{V}}_{2}$ are full rank matrices, the rank of the Hankel matrix ${\mathcal{H}}\left( {\bf{x}} \right)$ is indeed $P$, thereby indicating that ${\mathcal{H}}\left( {\bf{x}} \right)$ has low-rank property\cite{cai2019fast}. It is worth noting that a good choice for Hankel matrix size is $n_1  \approx n_2$\cite{chen2013spectral}. This ensures that the resulting matrix ${\mathcal{H}}\left( {\bf{x}} \right)$ is either a square matrix or an approximate square matrix. Specifically, in this paper, we adopt $n_1 = n_2 = \left( {\frac{{M + 1}}{2}} \right) $ if $M$ is odd, and $n_1 = n_2 - 1 = \left( {\frac{{M}}{2}} \right) $ if $M$ is even.

\begin{figure}
\centering
{\includegraphics[height=2.50 in]{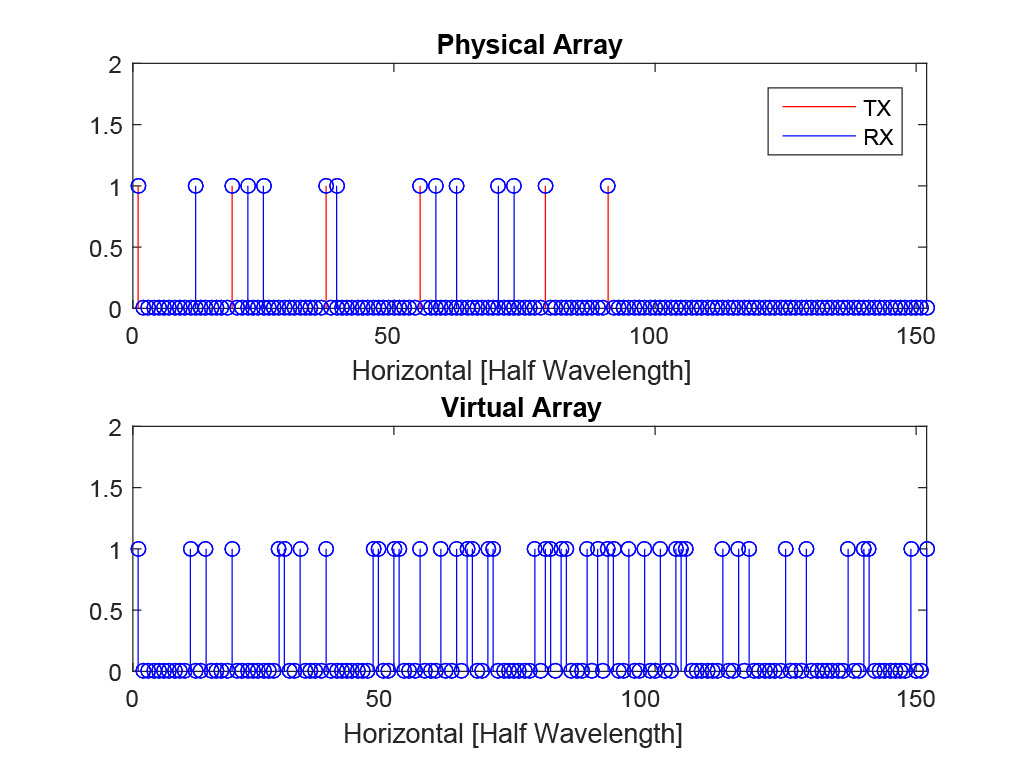}}
\caption{Example of an automotive radar with virtual sparse array of $48$ elements and aperture of $76\lambda$. 
}
\label{fig_1}
\end{figure}
We utilize a 1D virtual SLA synthesized by MIMO radar techniques \cite{sun2020mimo} with $M_{t}$ transmit antennas and $M_{r}$ receive antennas. The SLA has $M_{t}M_{r} < M$ elements while retaining the same aperture as ULA. Denote the array element indices of ULA as the complete set $\left \{ 1,2,\cdots,M \right \}$, the array element indices of SLA can be expressed as a subset ${\Omega^{\prime}}\subset\left \{ 1,2,\cdots ,M \right \}$. Thus, the signals received by the SLA can be viewed as partial observations of $\mathbf{x}$, and can be expressed as ${\bf{x}}_{s} = {\bf{m}}_{\Omega^{\prime}}\odot\mathbf{x}$, where ${{\bf{m}}_{\Omega^{\prime}}}= {\left[ {{m_1},{m_2}, \cdots ,{m_M}} \right]^{\top}}$ is a masking vector with $m_{j} = 1$, if $j\in\Omega^{\prime}$ or $m_{j} = 0$ if $j\notin\Omega^{\prime}$. Fig.~\ref{fig_1} shows an example of the array configuration of
an automotive radar which is a synthesized virtual sparse array with 48 elements of $1$ MIMO transceiver featuring 14 physical antennas,
where all transmit and receive antennas are clock synchronized. For the sake of simplicity in notation, hereafter, we denote the Hankel matrix of virtual sparse array response as $\mbX$.

\subsection{One-Bit Hankel Matrix Completion}
Assume $\Omega$ denotes the set of observed entries in $\mbX$ denoted by $\mathcal{P}_{\Omega}\left(\mbX\right)$, which is obtained by the array element indices of SLA $\Omega^{\prime}$. In one-bit sampled Hankel matrix completion, we solely observe the partial matrix through the $m^{\prime}$ one-bit samples, where $m^{\prime}\ll n_1 n_2$. The one-bit samples are generated by the following comparison between corresponding entries in $\mathcal{P}_{\Omega}\left(\mbX\right)$ and dither matrix $\boldsymbol{\mathcal{T}}=\left[\tau_{i,j}\right]\in \mathbb{C}^{n_1\times n_2}$ with $\tau_{i,j}=\tau^{(R)}_{i,j}+\mathrm{j} \tau^{(I)}_{i,j}$ and both real and imaginary parts follow $\mathcal{U}_{\left[-\frac{\Delta}{2},\frac{\Delta}{2}\right]}$:
\begin{equation}
\label{St_2}
\begin{aligned}
r^{(R)}_{i,j} &= \begin{cases} +1 & \operatorname{Re}\left(X_{i,j}\right)>\tau^{(R)}_{i,j},\\ -1 & \operatorname{Re}\left(X_{i,j}\right)<\tau^{(R)}_{i,j},
\end{cases} \quad (i,j)\in\Omega,
\end{aligned}
\end{equation}\normalsize
and
\begin{equation}
\label{St_2000}
\begin{aligned}
r^{(I)}_{i,j} &= \begin{cases} +1 & \operatorname{Im}\left(X_{i,j}\right)>\tau^{(I)}_{i,j},\\ -1 & \operatorname{Im}\left(X_{i,j}\right)<\tau^{(I)}_{i,j},
\end{cases} \quad (i,j)\in\Omega.
\end{aligned}
\end{equation}\normalsize
Therefore, the complex one-bit data is given by $r_{i,j} = r^{(R)}_{i,j}+\mathrm{j} r^{(I)}_{i,j}$. 
The acquired one-bit data forms the matrix $\mbR\in\left\{-1, 0, 1\right\}^{n_1\times n_2}+\mathrm{j} \left\{-1, 0, 1\right\}^{n_1\times n_2} $, where at the indices $(i,j)\in\Omega$, we have one-bit data, and the rest of the elements are zero.
In the work presented by the authors in \cite{candes2010matrix}, 
noisy matrix completion is formulated as a 
nuclear norm minimization problem. 
Consider the noisy measurements as follows:
\begin{equation}
X^{(n)}_{i,j} = X_{i,j}+Z_{i,j},\quad (i,j)\in \Omega,
\end{equation}
where $Z_{i,j}$ is a bounded additive noise. 

Extensive investigations conducted in \cite{candes2010matrix,cai2010singular} have demonstrated that matrix completion with noise can be formulated as a nuclear norm minimization problem as follows:
\begin{equation}
\label{Steph1}
\begin{aligned}
&\underset{\mbX}{\textrm{minimize}}\quad \left\|\mbX\right\|_{\star}\\
&\text{subject to} \quad \left\|\mathcal{P}_{\Omega}(\mbX-\mbX^{(n)})\right\|_{\mathrm{F}} \leq \delta,
\end{aligned}
\end{equation}
where $\mbX^{(n)}$ is the noisy matrix and $\delta$ presents the distortion effect. 
Let us assume that $\mathcal{Q}_{_{\Delta}}(\cdot)$ is the uniform quantizer with resolution parameter $\Delta\geq 0$ and $\mbQ = \mathcal{Q}_{_{\Delta}}\left(\operatorname{Re}\left(\mathcal{P}_{\Omega}\left(\mbX\right)\right)\right)+\mathrm{j}\mathcal{Q}_{_{\Delta}}\left(\operatorname{Im}\left(\mathcal{P}_{\Omega}\left(\mbX\right)\right)\right)$ represents the uniform quantization of known entries of low-rank matrix $\mbX$. Note that the quantizer is only applied to non-zero elements. Consequently, the quantized measurements can be expressed as follows:
\begin{equation}
\label{Steph2}
\mbQ = \mathcal{P}_{\Omega}\left(\mbX\right)+\mbN,
\end{equation}
where the matrix $\mbN\in\mathbb{C}^{n_1\times n_2}$ presents the effect of quantization as the additive noise matrix. Therefore, the nuclear norm minimization problem 
associated with the quantized MC is given by
\begin{equation}
\label{Steph3}
\begin{aligned}
&\underset{\mbX}{\textrm{minimize}}\quad \left\|\mbX\right\|_{\star}\\
&\text{subject to} \quad \left\|\mathcal{P}_{\Omega}(\mbX)-\mbQ\right\|_{\mathrm{F}} \leq \delta,
\end{aligned}
\end{equation}
where the parameter $\delta$ denotes the impact of the quantization process. 

In the literature \cite{thrampoulidis2020generalized}, it is well-known that the uniform quantizer effectively becomes a one-bit quantizer by limiting the measurement's dynamic range to half of the scale parameter of uniform dither. In other words, in matrix completion for both real and imaginary parts, we have
\begin{equation}
\mathcal{Q}_{_{\Delta}}\left(\mathcal{P}_{\Omega}\left(\mbX\right)\right) = \frac{\Delta}{2} \operatorname{sgn}\left(\mathcal{P}_{\Omega}\left(\mbX\right)-\mathcal{P}_{\Omega}\left(\boldsymbol{\mathcal{T}}\right)\right),~ \|\mbX\|_{\mathrm{max}}\leq \frac{\Delta}{2}.
\end{equation}
Hence, by fulfilling the dynamic range condition and employing the uniform dithering scheme, the quantized measurements $\mbQ$ can be substituted with $\frac{\Delta}{2}\mbR$. It is worth noting that in numerous applications, the upper bound of the dynamic range of measurements is known \cite{davenport20141,cai2013max}. Therefore, we can easily design the scale parameter of uniform dithers based on this information.

Denote a linear transformation $\mathcal{A}: \mathbb{C}^{n_1\times n_2}\rightarrow \mathbb{C}^{m^{\prime}}$ and $\mathcal{A}^{\star}:\mathbb{C}^{m^{\prime}}\rightarrow \mathbb{C}^{n_1\times n_2}$ as its adjoint operator. To address the nuclear norm minimization problem in one-bit Hankel matrix completion, we employ the SVT algorithm. If we consider the singular value decomposition (SVD) of $\mbX$ as $\mbX=\mbU\bSigma\mbV^{\top}$ and $\{\sigma_i\}$ as its singular values, the SVT use the singular value shrinkage operator comprehensively investigated in \cite{cai2010singular,ma2011fixed} which applies the partial SVD
to achieve the low-rank matrix structure as 
$\mathcal{D}_\tau(\mbX)=\mbU \mathcal{D}_\tau(\bSigma) \mbV^{\top},~ \mathcal{D}_\tau(\boldsymbol{\Sigma})=\operatorname{diag}\left(\left(\sigma_i-\tau\right)^{+}\right)$,
where $\tau\geq 0$ is the predefined threshold. The key distinction lies in our approach: rather than utilizing high-resolution partial measurements at each iteration, we employ one-bit data in the following manner:
\begin{equation}
\label{St_22}
\left\{\begin{array}{l}
\mbX^{(k)}=\mathcal{D}_\tau\left(\mathcal{A}^{\star}\left(\boldsymbol{y}^{(k-1)}\right)\right),\\
\boldsymbol{y}^{(k)}=\boldsymbol{y}^{(k-1)}+\delta_k \left(\mbb-\mathcal{A}\left(\mbX^{(k)}\right)\right),
\end{array}\right.
\end{equation} \normalsize
where $\left\{\delta_k\right\}$ are step sizes and $\mbb=\operatorname{vec}\left(\frac{\Delta}{2}[r_{i,j}]_{(i,j)\in \Omega}\right)\in\left\{-\frac{\Delta}{2},\frac{\Delta}{2}\right\}^{m^{\prime}}+\mathrm{j}\left\{-\frac{\Delta}{2},\frac{\Delta}{2}\right\}^{m^{\prime}}$. 
\subsection{Theoretical Guarantees for Recovery Performance}
As the norm-$2$ can be constrained by the norm-$1$, the one-bit matrix completion constraint in \eqref{Steph3} can be alternatively expressed as $\left\|\operatorname{vec}\left(\mathcal{P}_{\Omega}\left(\mbX\right)-\frac{\Delta}{2} \mbR\right)\right\|_{1}\leq q^{\prime}$. In the rest of this section, our aim is to theoretically investigate the parameter $q^{\prime}$ and establish its upper bound. Define the following operator:
\begin{definition}
\label{def_1}
For a matrix $\mbX=[X_{i,j}]\in\mathbb{R}^{n_1\times n_2}$ and the one-bit matrix $\mbR\in \left\{-1,0,1\right\}^{n_1\times n_2}$, we denote the average of distortions by
\begin{equation}
\label{a_3}
\begin{aligned}
T_{\mathrm{ave}}(\mbX) &= \frac{1}{m^{\prime}}\left\|\operatorname{vec}\left(\mathcal{P}_{\Omega}\left(\mbX\right)-\frac{\Delta}{2} \mbR\right)\right\|_{1}, \\&=\frac{1}{m^{\prime}}\sum_{(i,j)\in\Omega}\left|X_{i,j}-\frac{\Delta}{2}R_{i,j}\right|,
\end{aligned}
\end{equation}
where $m^{\prime}=|\Omega|$. 
\end{definition}
It is important to note that the guarantee is obtained under the uniform dithering scheme. In the following definition, we state the consistent reconstruction property which will be our assumption in the provided theorem:
\begin{definition}
\label{def_2}
Define a low-rank matrix $\mbX=[X_{i,j}]\in\mathbb{R}^{n_1\times n_2}$. The consistency property of uniform quantization over the pair $\mbX,\mbY\in\mathcal{K}_r$, is given by
\begin{equation}
\label{a_1}
\mathcal{Q}_{\Delta}\left(\mathcal{P}_{\Omega}\left(\mbX\right)\right)=\mathcal{Q}_{\Delta}\left(\mathcal{P}_{\Omega}\left(\mbY\right)\right).
\end{equation}
\end{definition}
When we frame the one-bit matrix completion problem as a nuclear norm minimization problem with the one-bit data matrix $\mbR\in\left\{-1,0,1\right\}^{n_1\times n_2}+\mathrm{j}\left\{-1,0,1\right\}^{n_1\times n_2}$ representing our observation, we can derive an upper bound for the recovery, as presented in the following theorem:
\begin{theorem}
\label{Theorem_q}
Define the set $\mathcal{K}_r$ as
\begin{equation}
\label{n_02}
\mathcal{K}_r=\left\{\mbX^{\prime}\in\mathbb{C}^{n_1\times n_2}\mid\operatorname{rank}(\mbX^{\prime})\leq r, \beta\leq\alpha\right\}\subset\mathbb{C}^{n_1\times n_2},
\end{equation}
where $\beta= \max\left(\|\operatorname{Re}\left(\mbX^{\prime}\right)\|_{\mathrm{max}},\|\operatorname{Im}\left(\mbX^{\prime}\right)\|_{\mathrm{max}}\right)$. Consider a matrix $\mbX\in\mathcal{K}_r$. Now, let's assume that $m^{\prime}$ entries of $\mbX$, randomly selected using uniform sampling, undergo dithered one-bit quantization with uniform thresholds generated as $\tau_{i,j}\sim \mathcal{U}_{\left[-\frac{\Delta}{2},\frac{\Delta}{2}\right]}$ where $2\alpha\leq \Delta$, leading to the observed one-bit data matrix $\mbR$. 
Then, with a probability exceeding $1-4e^{-\frac{\varepsilon^2 m^{\prime}}{\Delta^2}}$, we can assert that the recovery error between $\mbX$ and the estimated matrix $\bar{\mbX}$ satisfying the consistency in Definition~\ref{def_2} is bounded as follows:
\begin{equation}
\label{a_50000}
\|\mbX-\bar{\mbX}\|_{\mathrm{F}}\leq2\sqrt{8\epsilon\alpha n_1n_2},
\end{equation}
as long as $m^{\prime} \gtrsim \varepsilon^{-\frac{5}{2}} r \max\left(n_1,n_2\right)$.
\end{theorem}
The complete proof of Theorem~\ref{Theorem_q} can be found in Appendix~\ref{proof0}. According to Theorem~\ref{Theorem_q}, the recovery performance of dithered one-bit matrix completion depends on the dimension of signal and the distortion of uniform quantized values.
\begin{figure}
	\centering
	{\includegraphics[height=2.50 in]{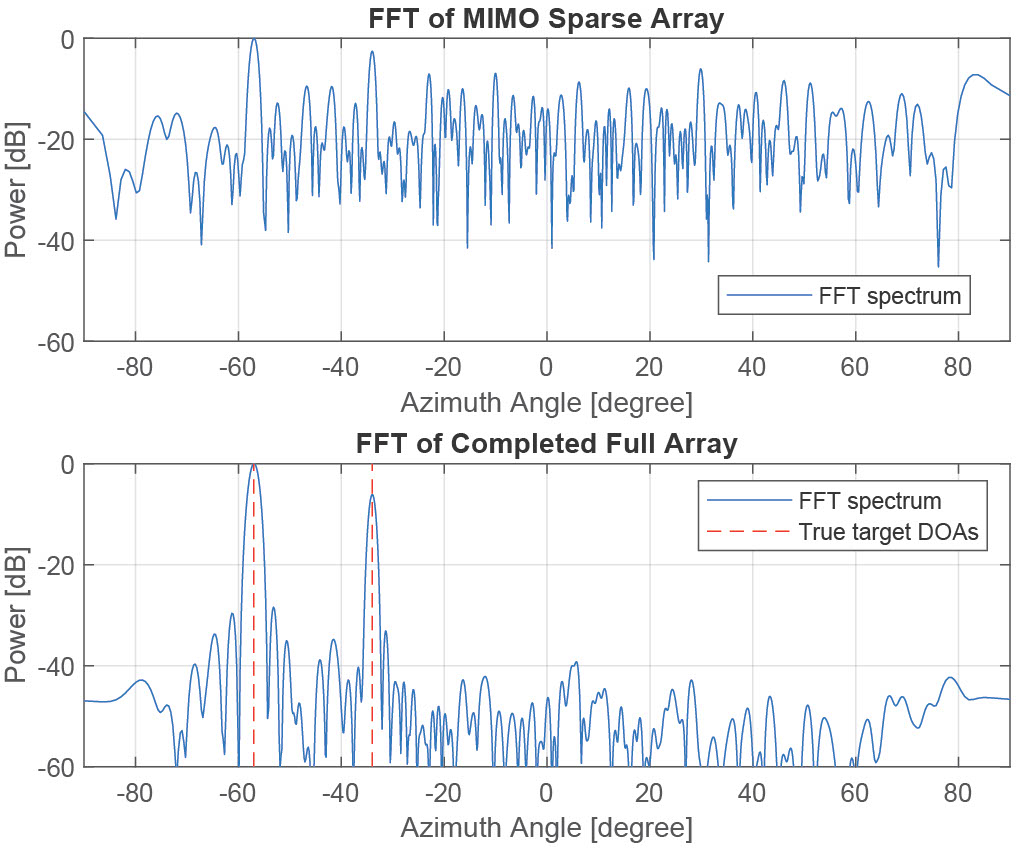}}
	\caption{The spectrum of two targets with azimuth angles of $\theta_1 = -57^{\circ} $ and
		$\theta_2 = -34^{\circ}$ under MIMO sparse array and fully completed array. 
	}
	\label{fig_2}
	\vspace{-5pt}
\end{figure}
\section{Numerical Investigation}
In this section, we undertake numerical investigations to assess the performance of the one-dimensional sparse array completion that jointly utilizes sparse spectrum and sparse arrays with one-bit dither measurements for radar sensing in automotive applications. 

To attain high azimuth angular resolution, we cascade multiple automotive radar transceivers to synthesize a large sparse array in azimuth. We focus on the same physical array illustrated in Fig.~\ref{fig_1}, where there are $M_t=6$ transmit antennas and $M_r=8$ receive antennas arranged in an interleaved manner along the horizontal direction at:
\begin{equation}
\begin{aligned}
& l_{\mathrm{TX}}=[1,19,37,55,79,91] \lambda / 2, \\
& l_{\mathrm{RX}}=[12,22,25,39,58,62,70,73] \lambda / 2.
\end{aligned}
\end{equation}
A virtual array with total $48$ elements is synthesized. The transmit and receive antennas as well as the virtual array are plotted
in Fig.~\ref{fig_1}. 
Two targets are positioned at the same range $R = 100~\text{m}$ with velocities of $v = -10~\text{m/s}$. Their respective azimuth angles are $\theta_1 = -57^{\circ}$ and $\theta_2 = -34^{\circ}$. The SNR is set as $20$dB in our simulations. Initially, the two targets can be first separated in range-Doppler. The complex peak values in the range-Doppler spectrum, corresponding to each virtual sparse array, constitute an array snapshot for azimuth angle determination. We apply the one-bit quantizer with uniform dithering following $\tau_{i,j}\sim \mathcal{U}_{\left[-\frac{\Delta}{2},\frac{\Delta}{2}\right]}$, to a Hankel matrix $\mbX\in\mathbb{C}^{76\times 76}$, which is constructed based on the array response of a ULA with
$152$ elements and sampled by the SLA shown in Fig.~\ref{fig_1}. The array response of the SLA is normalized
by its first element. The scale parameter of uniform dithers is designed in such a way to dominate the dynamic range of measurements. Using the partial one-bit data $\mbR$, the rank-$2$
Hankel matrix $\mbX$ is completed via the proposed algorithm in \eqref{St_22}. Let
$\bar{\mbX}$ denote the completed Hankel matrix. The full ULA response
can be reconstructed by taking the average of the anti-diagonal elements of matrix $\bar{\mbX}$. The completed full array has an aperture size of $76\lambda$. 

In Fig.~\ref{fig_2}, we depict the angle spectrum for the two targets. The azimuth angle spectra are derived by applying FFT to the original SLA with the holes filled with zeros and to the full array completed via one-bit matrix completion, respectively. The FFT of the SLA produces two peaks corresponding to the correct azimuth directions but with high sidelobes, making it challenging to detect the two targets accurately in azimuth. In contrast, the completed full array from one-bit samples exhibits two distinct peaks corresponding to the correct azimuth locations in the angle spectrum, with significantly suppressed sidelobes.
\section{Discussion}
We employed the memoryless scalar quantization to coarsely quantize measurements from a sparse linear array system used in the high-resolution imaging radar. By utilizing a uniform dithering scheme and recording only one-bit data for reconstruction, we accurately detected two targets in azimuth with significantly suppressed spectrum sidelobes. The quantized Hankel matrix completion problem was solved, and we extensively discussed theoretical guarantees and recovery performance. This paper specifically addressed the two-dimensional scenario, and 
future work would aim to extend the scope to three-dimensional targets, incorporating enriched Hankel matrix completion techniques for a more comprehensive analysis.

\appendices
\section{Proof of Theorem~\ref{Theorem_q}}
\label{proof0}
Without loss of generality, we assume $\alpha=\frac{\Delta}{2}$. We begin the proof by presenting the following lemma:
\begin{lemma}
\label{lem_1}
In the settings of Definition~\ref{def_1}, we have
\begin{equation}
\label{a_6}
\operatorname{Pr}\left(\sup_{\mbX\in\mathcal{K}_r}\left|T_{\mathrm{ave}}(\mbX)-\alpha+\frac{\|\mbX\|_{\mathrm{F}}^2}{\alpha n_1n_2}\right|\geq\epsilon\right)\leq 2e^{-\frac{\epsilon^2 m^{\prime}}{4\alpha^2}},
\end{equation}
where $\epsilon$ is a positive value.
\end{lemma}
\begin{IEEEproof}
For simplicity of notation, denote $d_{i,j}=\left|X_{i,j}-\alpha R_{i,j}\right|$. Then, we can write
\par\noindent\small
\begin{equation}
\label{a_29}
\begin{aligned}
\mathbb{E}_{\tau}\left\{d_{i,j}\right\}&=\frac{1}{2\alpha}\int_{-\alpha}^{\alpha}\left|X_{i,j}-\alpha R_{i,j}\right|\,d\tau\\&=\frac{1}{2\alpha}\left[\int_{-\alpha}^{X}\alpha-X_{i,j}\,d\tau+\int_{X}^{\alpha}\alpha+X_{i,j}\,d\tau\right]\\&=\alpha-\frac{X^2_{i,j}}{\alpha}.
\end{aligned}
\end{equation}\normalsize
Therefore, we have 
\begin{equation}
\label{a_30}
\begin{aligned}
\mathbb{E}_{\tau}\left\{T_{\mathrm{ave}}(\mbX)\right\}&=\frac{1}{m^{\prime}}\sum_{(i,j)\in\Omega}\left(\alpha-\frac{X_{i,j}^2}{\alpha}\right)\\&=\alpha-\frac{\|\mathcal{P}_{\Omega}\left(\mbX\right)\|_{\mathrm{F}}^2}{\alpha m^{\prime}}.
\end{aligned}
\end{equation}
Computing the expected value of \eqref{a_30} respect to the randomness of $(i,j)$ leads to 
\begin{equation}
\label{ghombol}
\mathbb{E}_{\tau,(i,j)}\left\{T_{\mathrm{ave}}(\mbX)\right\}=\alpha-\frac{\|\mbX\|_{\mathrm{F}}^2}{\alpha n_1n_2}.
\end{equation} 
In the following lemma, we present the Hoeffding's inequality for bounded random variables:
\begin{lemma}\cite[Theroem~2.2.5]{vershynin2018high}
\label{lemma_hoeffding} 
Let $\left\{X_i\right\}^{n}_{i=1}$ be independent, bounded random variables
satisfying $X_i \in [a_i, b_i]$, then for any $t > 0$ it holds that
\begin{equation}
\operatorname{Pr}\left(\left|\frac{1}{n} \sum_{i=1}^n\left(X_i-\mathbb{E}\left\{ X_i\right\}\right)\right| \geq t\right)\leq 2 e^{-\frac{2 n^2 t^2}{\sum_{i=1}^n\left(b_i-a_i\right)^2}}.  
\end{equation}
\end{lemma}
Note that for each random variable $d_{i,j}$, we have $0\leq d_{i,j}\leq 2\alpha$. Then, following Lemma~\ref{lemma_hoeffding}, we can write
\begin{equation}
\label{a_31}
\operatorname{Pr}\left(\left|T_{\mathrm{ave}}(\mbX)-\alpha+\frac{\|\mbX\|_{\mathrm{F}}^2}{\alpha n_1n_2}\right|\geq\epsilon\right)\leq 2e^{-\frac{\epsilon^2 m^{\prime}}{2\alpha^2}}.
\end{equation}
As we consider the supremum over all $\mbX\in\mathcal{K}_r$, it is necessary to multiply the resulting probability by the covering number of the defined set. It is straightforward to verify that the covering number of $\rho$-balls required to cover the set $\mathcal{K}_{r}$ is upper bounded by
\begin{equation}
\mathcal{N}\left(\mathcal{K}_{r}, \left\|\cdot\right\|_{\mathrm{F}}, \rho\right)\leq \left(1+\frac{2\alpha\sqrt{n_1n_2}}{\rho}\right)^{(n_1+n_2)r},
\end{equation}
which can be further upper bounded by
\begin{equation}
\label{cov}
\begin{aligned}
\mathcal{N}\left(\mathcal{K}_{r}, \left\|\cdot\right\|_{\mathrm{F}},\rho\right)&\leq e^{(n_1+n_2)r\log\left(1+\frac{2\alpha\sqrt{n_1n_2}}{\rho}\right)}\\&\leq e^{\frac{2\alpha(n_1+n_2)r\sqrt{n_1n_2}}{\rho}}.
\end{aligned} 
\end{equation} 
Based on \eqref{cov}, the Kolmogorov $\rho$-entropy of the set $\mathcal{K}_r$ is upper bounded by
\begin{equation}
\label{cov23}
\begin{aligned}
\mathcal{H}\left(\mathcal{K}_{r},\rho\right)\leq \frac{2\alpha(n_1+n_2)r\sqrt{n_1n_2}}{\rho}.
\end{aligned} 
\end{equation}
To achieve the probability at least $1-2e^{-\frac{\epsilon^2 m^{\prime}}{4\alpha^2}}$, it is sufficient to write
\begin{equation}
e^{\frac{2\alpha (n_1+n_2)r\sqrt{n_1n_2}}{\rho}}\leq e^{\frac{\epsilon^2 m^{\prime}}{4\alpha^2}}, 
\end{equation}
or equivalently,  
\begin{equation}
m^{\prime}\geq \frac{8\alpha^{3}(n_1+n_2)r\sqrt{n_1n_2}}{\epsilon^2\rho}, 
\end{equation}
which proves the lemma.
\end{IEEEproof}
Define $\mbZ=\frac{1}{2}(\mbX+\bar{\mbX})$. We can write $Z_{i,j}-\alpha R_{i,j}=\frac{1}{2}\left(X_{i,j}-\alpha R_{i,j}+\bar{X}_{i,j}-\alpha R_{i,j}\right)$. The triangle inequality implies
\begin{equation}
\label{a_33}
\left|Z_{i,j}-\alpha R_{i,j}\right|\leq\frac{1}{2}\left(\left|X_{i,j}-\alpha R_{i,j}\right|+\left|\bar{X}_{i,j}-\alpha R_{i,j}\right|\right).
\end{equation}
Based on \eqref{a_33}, for all $(i,j)\in\Omega$, we can write
\begin{equation}
\label{a_34}
T_{\mathrm{ave}}(\mbZ)\leq\frac{1}{2}\left[T_{\mathrm{ave}}(\mbX)+T_{\mathrm{ave}}(\bar{\mbX})\right].
\end{equation}
Under the consistent reconstruction assumption in Definition~\ref{def_2}, with a failure probability at most $2e^{-\frac{\epsilon^2 m^{\prime}}{4\alpha^2}}$, Lemma~\ref{lem_1} implies
\begin{equation}
\label{a_new}
\left|T_{\mathrm{ave}}(\bar{\mbX})-\alpha+\frac{\|\bar{\mbX}\|_{\mathrm{F}}^2}{\alpha n_1n_2}\right|\leq\epsilon.
\end{equation}
Similarly, with a failure probability at most $2e^{-\frac{\epsilon^2 m^{\prime}}{4\alpha^2}}$, we have
\begin{equation}
\label{a_9}
\frac{\|\mbZ\|_{\mathrm{F}}^2}{\alpha n_1n_2}\geq \alpha-\epsilon-T_{\mathrm{ave}}(\mbZ),
\end{equation}
which together with \eqref{a_34} and \eqref{a_new}, we can write
\par\noindent\small
\begin{equation}
\label{a_10}
\begin{aligned}
\frac{\|\mbZ\|_{\mathrm{F}}^2}{\alpha n_1n_2}&\geq\alpha-\epsilon-\frac{1}{2}\left[T_{\mathrm{ave}}(\mbX)+T_{\mathrm{ave}}(\bar{\mbX})\right]\\&\geq\alpha-\epsilon-\frac{1}{2}\left[-\frac{\|\mbX\|_{\mathrm{F}}^2}{\alpha n_1n_2}+\alpha+\epsilon-\frac{\|\bar{\mbX}\|_{\mathrm{F}}^2}{\alpha n_1n_2}+\alpha+\epsilon\right]\\&=\frac{1}{2\alpha n_1n_2}\left[\|\mbX\|_{\mathrm{F}}^2+\|\bar{\mbX}\|_{\mathrm{F}}^2\right]-2\epsilon.
\end{aligned}
\end{equation}\normalsize
Based on the definition of $\mbZ$, we can rewrite \eqref{a_10} in terms of $\mbX$ and $\bar{\mbX}$ as follows
\begin{equation}
\label{a_11}
\|\mbX+\bar{\mbX}\|_{\mathrm{F}}^2\geq2\left(\|\mbX\|_{\mathrm{F}}^2+\|\bar{\mbX}\|_{\mathrm{F}}^2\right)-8\alpha n_1n_2\epsilon.
\end{equation}
By the parallelogram law, we conclude that
\begin{equation}
\label{a_12}
\begin{aligned}
\|\mbX-\bar{\mbX}\|_{\mathrm{F}}^2&=2\left[\|\mbX\|_{\mathrm{F}}^2+\|\bar{\mbX}\|_{\mathrm{F}}^2\right]-\|\mbX+\bar{\mbX}\|_{\mathrm{F}}^2\\&\leq8\alpha n_1n_2\epsilon.
\end{aligned}
\end{equation}
Denote $\rho=\sqrt{8\epsilon\alpha n_1n_2}$. Then, according to Lemma~\ref{lem_1}, with a failure probability at most $2e^{-\frac{\epsilon^2 m^{\prime}}{4\alpha^2}}$, we have $\|\mbX-\bar{\mbX}\|_{\mathrm{F}}\leq\rho$ as long as 
\begin{equation}
m^{\prime}\gtrsim \epsilon^{-\frac{5}{2}}r\max\left(n_1,n_2\right).
\end{equation}
To finalize the proof, we now take into account both real and imaginary parts. It is noteworthy that the error of a complex value can be upper bounded by its real part and imaginary part, based on the triangle inequality, as follows:
\begin{equation}
\|\mbX-\bar{\mbX}\|_{\mathrm{F}} \leq \|\operatorname{Re}\left(\mbX\right)-\operatorname{Re}\left(\bar{\mbX}\right)\|_{\mathrm{F}}+\|\operatorname{Im}\left(\mbX\right)-\operatorname{Im}\left(\bar{\mbX}\right)\|_{\mathrm{F}}, 
\end{equation}
where both parts are upper bounded by \eqref{a_12} leading to have
\begin{equation}
\|\mbX-\bar{\mbX}\|_{\mathrm{F}} \leq 2\rho, 
\end{equation}
which completes the proof. Note that the factor of $4$ in the probability arises from the fact that both real and imaginary parts have the same probability. According to the union bound, this results in the overall probability being twice as large.
\bibliographystyle{IEEEtran}
\bibliography{references}

\end{document}